\documentclass[12pt,a4paper,twoside]{article}
\newenvironment{changemargin}[2]{%
\begin{list}{}{%
\setlength{\leftmargin}{#1}%
\setlength{\rightmargin}{#2}%
}%
\item[]}
{\end{list}}
\usepackage{graphicx}
\usepackage{color}
\usepackage{lscape}
\usepackage{hyperref}
\usepackage{epstopdf}
\usepackage{lscape}
\textheight = 9.20in\usepackage{amsmath}
\usepackage{hyperref}
\usepackage{amsmath}
\usepackage{setspace}
\usepackage[round, sort]{natbib}

\begin{document}
\baselineskip=0.25in
{\bf \LARGE
\begin{changemargin}{-1.2cm}{0.5cm}
\begin{center}
Triangular libration points in the R3BP under combined effects of oblateness, radiation and power-law profile
\end{center}
\end{changemargin}}
\vspace{2mm}
\begin{center}
{\large{\bf B. J. Falaye $^a$$^{,}$$^b$$^{,}$$^\dag$$^{,}$}}\footnote{\scriptsize E-mail:~ fbjames11@physicist.net;~ babatunde.falaye@fulafia.edu.ng\\ \dag Corresponding Author Tel. no: (+52) 1 5521450362}\large{\bf ,} {\large{\bf Shi Hai Dong $^c$$^{,}$}}\footnote{\scriptsize E-mail:~ dongsh2@yahoo.com}\large{\bf ,} {\large{\bf K. J. Oyewumi $^d$$^{,}$}}\footnote{\scriptsize E-mail:~ kjoyewumi66@gmail.com}\large{\bf ,} {\large{\bf O. A. Falaiye $^b$$^\ddag$$^{,}$}}\footnote{\scriptsize E-mail:~ sesantayo2001@yahoo.com\\ \ddag On sabbatical leave from University of Ilorin, Ilorin, Nigeria}\large{\bf ,} {\large{\bf E. S. Joshua $^b$$^{,}$}}\footnote{\scriptsize E-mail:~ dreamsambo@yahoo.com}\large{\bf ,} {\large{\bf J. Omojola $^b$$^{,}$}}\footnote{\scriptsize E-mail:~ friendsaver1@yahoo.com}\large{\bf ,} {\large{\bf O. J. Abimbola $^b$$^{,}$}}\footnote{\scriptsize E-mail:~ ladiran@gmail.com}\large{\bf ,} {\large{\bf O. Kalu $^b$$^{,}$}}\footnote{\scriptsize E-mail:~ konyekachi@gmail.com} \large{\bf and} {\large{\bf S. M. Ikhdair $^e$$^{,}$}}\footnote{\scriptsize E-mail:~ sameer.ikhdair@najah.edu;~ sikhdair@gmail.com.}
\end{center}
{\footnotesize
\begin{center}
{\it $^\textbf{a}$Departamento de F\'isica, Escuela Superior de F\'isica y Matem\'aticas, Instituto Polit\'ecnico Nacional, Edificio 9, Unidad Profesional Adolfo L\'opez Mateos, Mexico D.F. 07738, Mexico.}\\
{\it $^\textbf{b}$Astrophysics Group, Department of Physics, Federal University Lafia,  P. M. B. 146, Lafia, Nigeria.}
{\it $^\textbf{c}$CIDETEC, Instituto Polit\'{e}cnico Nacional, UPALM, M\'{e}xico D. F. 07700, M\'{e}xico.}
{\it $^\textbf{d}$Theoretical Physics Section, Department of Physics, University of Ilorin, P. M. B. 1515, Ilorin, Nigeria.}
{\it $^\textbf{e}$Department of Physics, Faculty of Science, An-Najah National University, New campus, Nablus, West Bank, Palestine.}
\end{center}}
\textbf{
\begin{center}
Accepted for publication: Advances in Space Research (2015)
\end{center}
}
\begin{abstract}
\noindent
We study the effects of oblateness up to $J_4$ of the primaries and  power-law density profile (PDP) on the linear stability of libration location of an infinitesimal mass within the framework of restricted three body problem (R3BP), by using a more realistic model in which a disc with PDP is rotating around the common center of the system mass with perturbed mean motion. The existence and stability of triangular equilibrium points have been explored. It has been shown that triangular equilibrium points are stable for $0<\mu<\mu_c$ and unstable for $\mu_c\leq\mu\leq1/2$, where $\mu_c$ denotes the critical mass parameter. We find that, the oblateness up to $J_2$ of the primaries and the radiation reduces the stability range while the oblateness up to $J_4$ of the primaries increases the size of stability both in the context where PDP is considered and ignored. The PDP has an effect of about $\approx0.01$ reduction on the application of $\mu_c$ to Earth-Moon and Jupiter-Moons systems. We find that the comprehensive effects of the perturbations have a stabilizing proclivity. However, the oblateness up to $J_2$ of the primaries and the radiation of the primaries have tendency for instability, while coefficients up to $J_4$ of the primaries have stability predisposition. In the limiting case $c=0$, and also by setting appropriate parameter(s) to zero, our results are in excellent agreement with the ones obtained previously. Libration points play a very important role in space mission and as a consequence, our results have a practical application in space dynamics and related areas. The model may be applied to study the  navigation and stationkeeping operations of spacecraft (infinitesimal mass) around the Jupiter (more massive) -Callisto (less massive) system,  where PDP accounts for the circumsolar ring of asteroidal dust, which has a cloud of dust permanently in its wake.
\end{abstract}
{\bf Keywords}: Restricted three-body problem; Libration points; Oblateness; Power-law density profile.

\section{Introduction}
Three-body problem is an important class of problem in classical and quantum mechanics which involves modeling the motion of three particles or massive bodies subject to their mutually perturbing gravitational attractions, e.g., the Sun-Earth-Moon system. For this system, the Sun's mass is so dominant that it can be treated as a fixed object and the Earth-Moon system is treated as a two-body system from the point of view of a reference frame orbiting the Sun with that system. The simplest form of the three-body problem is the restricted three-body problem (R3BP) \citep{YE1}. It describes the motion of an infinitesimal mass moving under the gravitational influence of two massive bodies known as the primaries which move in a circular orbits around their center of mass on the account of their mutual attractions. Moreover, the infinitesimal mass is not influenced by the motion of the primaries. 

To tackle the problem of classical restricted three-body problem, Lagrange considered the behavior of the distances between the bodies without finding a general solution. He obtained numerous equations from which he discovered two distinct classes of constant-pattern solutions. The first is referred to as the collinear (in which one of the distances is the sum of the other two) and the second is equiangular (in which the three distances are equal). The two distinct classes result into Lagrangian points $L_{1}, L_2, L_3, L_{4}$ and $L_5$, respectively which form the five special points in rotating frame of reference where gravitational equilibrium can be maintained. 

The triangular libration points of the dynamical system denoted by $L_{4}$ and $L_{5}$ are sometimes called as the triangular Lagrange points or Trojan points. The name ``Trojan points" comes from the Jupiter Trojans at the Sun-Jupiter $L_4$ and $L_5$ points, named after characters from Homer's Iliad. The Sun's gravity perturbs objects in the libration points so as not to be a perfect picture, but rather fairly close. The distances to $m_1$ and $m_2$ are equal at $L_{4,5}$ and thus, make the points balance. The resultant force acts through the barycenter of the system due to the gravitational forces from the two massive bodies which are in the same ratio as the masses of the primaries. 

\begin{figure}[ht]
 \includegraphics[height=85mm,width=120mm]{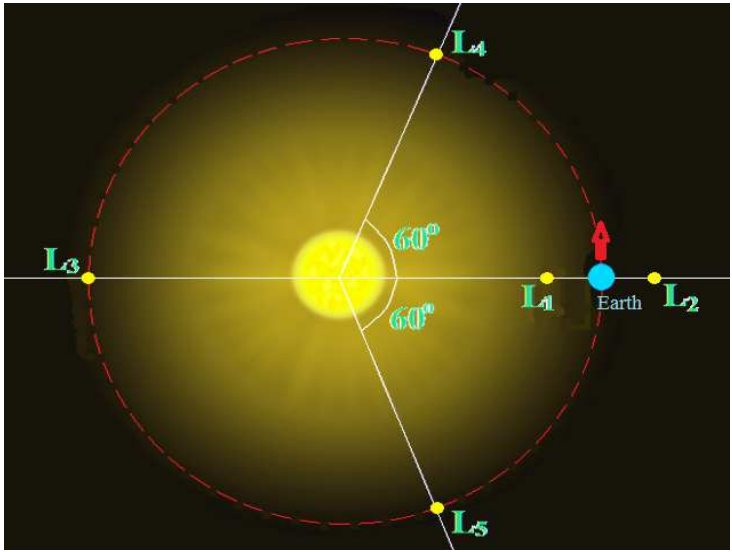}
\caption{(Color online) Position of Lagrangian points in a Sun-Earth system}
\label{fig1}
\end{figure}

In Figure \ref{fig1}, we show the position of these  Lagrangian points in a Sun-Earth system. These five Lagrangian points have very important applications in astronautical. A great number of space missions have been completed while some operations are still in progress. The Solar and Heliospheric Observatory (SOHO)  lunched in 1995 and Microwave Anisotropy Probe (MAP) lunched in 2001 by NASA are currently in operation Sun-Earth $L_1$ and $L_2$ respectively. Solar TErrestrial RElations Observatory-Ahead (STEREO-A) made its closest pass to $L_5$ recently, on its orbit around the Sun. Asteroid 2010 SO16, is currently proximal to $L_5$ but at a high inclination. In view of the importance of Lagrange points to the exploration and development of space, the dynamics and stability of a satellite were studied under multiple Trojan asteroids influence in ref. \citep{YE2}.

In studying Classical R3BP, it is required to put into consideration the oblateness of a body because, some bodies in our solar system for instance the earth and some in stellar system such as Altair are not spherical but oblate. In light of this, an enormous number of studies have been made in the recent years. The literature is huge, however, the following references give some older as well as recent studies.  Letelier and Silva obtained a particular solution to the restricted three-body problem where the bodies are allowed to either lose or gain mass to or from a static atmosphere \citep{YE3}. The motion of an infinitesimal body in the generalized restricted three-body problem have been presented by Singh and Taura \citep{YE4}. It was found that, in addition to the usual five equilibrium points, there appear two new collinear points $L_{n1}$, $L_{n2}$ due to the potential from the belt, and in the presence of all these perturbations, the equilibrium points $L_1$, $L_3$ come nearer to the primaries; while $L_2$, $L_4$, $L_5$, $L_{n1}$ move towards the less massive primary and $L_{n2}$ moves away from it.

The dynamics of galactic systems with central binary black holes was studied by modifying the restricted three body problem, in which a galactic potential was added as an external potential \citep{YE5}. Narayan and Shrivastava \citep{YE6} discussed the effects of oblateness and the photo-gravitational of the primaries on the location and the stability of the triangular equilibrium points in the elliptical restricted three-body problem. Abouelmagd and Sharaf \citep{YE7} explored the existence of libration points and their linear stability when the more massive primary is radiating and the less massive primary is an oblate spheroid. The existence and linear stability of equilibrium points in the perturbed Robe's circular restricted three-body problem under the assumption that the hydrostatic equilibrium figure of the first primary is an oblate spheroid has been examined recently \citep{YE8}. 

Analytical study of the dynamics of the third body within the framework of R3PB having a variable mass which changes accordingly
to Jeansa law has been presented recently by Abouelmagd and Mostafa \citep{YE9}. In ref. \citep{YE10}, the author investigated the stability of triangular libration points in a planar restricted elliptic three-body problem for two sets of parameters corresponding to the cases of double resonances. The existence of the libration points and their linear stability within the framework of the R3PB, considering the effects of the first two even zonal harmonics parameters with respect to both primaries have been studied recently \citep{YE11}. Couple R3BP is one of the examples of restricted four-body problem (R4BP). Within this context, Falaye \citep{YE12} studied the motion of an infinitesimal mass by assuming that the primaries of the system are radiating-oblate spheroids surrounded by a circular cluster of material points.

In this research work, we intend to extend the work of Singh and Taura \citep{YE1} by considering a modified restricted three body problem without the effect from circular cluster of material point but with a disk, which rotates around the center of mass of the system with perturbed mean motion and also, the work of Kushvah and Kishor \citep{YE13} by examine the effect of oblateness of the primaries up to $J_4$ of the zonal harmonics on the stability of equilibrium points. Our results will also generalize the recent work \citep{YE11}.

\section{Mathematical formulation of the model}
Let $m_i$ $(i=1,2)$ denote the masses of the primaries ($m_1$ for the more massive primary and $m_2$ for the less massive primary) and let the mass of an infinitesimal body moving in the plane of motion of the primaries be $m$. The positions of the primaries are defined relative to a rotating coordinate frame $oxyz$ whose $x$-axis coincides with the line joining $m_1$ and $m_2$ and whose origin coincides with the center of mass of $m_1$ and $m_2$; while $y$-axis is perpendicular to the $x$-axis, the $z$-axis is perpendicular to the orbital plane of the primaries. Let $r_1$ and $r_2$ be the distances between $m$ and the primaries $m_1$ and $m_2$ respectively, and $R$ be the distance between the primaries. $m$ moves in orbital plane under their mutual gravitational fields. The sum of $m_1$ and $m_2$ is unity and $\mu=m_2/(m_1+m_2)$ is the mass ratio with the distance between taken as one \citep{YE14}. Furthermore, the unit of time is chosen to make both the constant of gravitation and the unperturbed mean motion equal to $1$. To study the position of the infinitesimal mass in the plane of motion of the primaries, we can either use the sidereal system of coordinates, or the synodical system of coordinates. In a synodical system of coordinates, we write the coordinates of $m_1$, $m_2$ and $m$ as $(-\mu, 0, 0)$, $(1-\mu, 0, 0)$ and $(x, y, z)$ respectively. Hence, the kinetic energy (K.E.) of the infinitesimal mass in this barycentric coordinate system rotating about $z$-axis with constant angular velocity $n$, is represented as
\begin{equation}
K.E.=\frac{1}{2}m\left[n^2(x^2+y^2)+2n(x\dot{y}-y\dot{x})+(\dot{x}^2+\dot{y}^2)\right],
\label{E1}
\end{equation}
where the dot denotes differential with respect to time.  Now, we take the potential energy of the infinitesimal mass under radiating-oblate primaries as 
\begin{equation}
V=-Gm\left[q_1(1-\mu)\left(\frac{1}{r_1}+\frac{A_1}{2r_1^3}-\frac{3A_2}{8r_1^5}\right)+q_2\mu \left(\frac{1}{r_2}+\frac{B_1}{2r_2^3}-\frac{3B_2}{8r_2^5}\right)\right].
\label{E2}
\end{equation}
Furthermore, we consider a more realistic model in which a disk with PDP, is rotating around the common center of mass of the system. The PDP of the disc having a thickness $h\approx 10^{-4}$ is $\rho(r)=cr^{-p}$  \citep{YE13}, where $p$ denotes a natural number which we take as $3$ and $c$ is a constant determined with the help of the disc mass. Thus, equation (\ref{E2}) becomes:
\begin{equation}
V=-Gm\left[q_1(1-\mu)\left(\frac{1}{r_1}+\frac{A_1}{2r_1^3}-\frac{3A_2}{8r_1^5}\right)+q_2\mu \left(\frac{1}{r_2}+\frac{B_1}{2r_2^3}-\frac{3B_2}{8r_2^5}\right)\right]-V_{j,k},
\label{E3}
\end{equation}
with $r_i(i=1,2)$ define as
\begin{equation}
r_1=\sqrt{(x+\mu)^2+y^2},\ \ \ r_2=\sqrt{(x+\mu-1)^2+y^2},
\label{E4}
\end{equation}
where $G$ is the gravitational constant. $q_1$ and $q_2$ are the radiation factors of the  primaries. Parameter $q_i = 1-\frac{F_{r_i}}{F_{g_i}}$ $(i=1,2)$, where $F_r$ is the force caused by radiation pressure force and $F_g$ results due to gravitational force \citep{YE1}. The oblateness coefficients for the bigger primary is denoted as $A_i$, with $0<A_i =J_{2i}R_1^{2i}<<1$ and for the smaller primary as $B_i$, with $0<B_i=J_{2i}R_1^{2i}<<1$, $i=1,2$, where $J_{2i}$ are zonal harmonic coefficients and $R_{1,2}$ denote the mean radii of $m_{1,2}$, supposing the primaries have their equatorial planes coinciding with the plane of motion. $V_{j,k}$ is the potential due to the disc which is given by \citep{YE13,YE06}
\begin{equation}
 V_{j,k}=-4\int_{r'}\frac{F(\zeta)\rho'(r')r'}{r+r'}dr',
\label{E6}
\end{equation}
where $F(\zeta)$ denotes the elliptic integral of the first kind, $r'$ is the disk's reference radius and $\zeta={2\sqrt{rr'}}/{(r+r')}$. We write the Lagrangian to our problem as
\begin{equation}
 L=\frac{mn^2}{2}\left(x^2+y^2\right)+mn\left(x\dot{y}-\dot{x}y\right)+\frac{m}{2}\left(\dot{x}^2+\dot{y}^2\right)-V_{j,k},
\label{E7}
\end{equation}
where the mean motion $n$ is 
\begin{equation}
 n^2=1+\frac{3}{2}\left(A_1+B_1-\frac{5}{4}(A_2+B_2)\right)-\left.2f_b(r)\right|_{r=r_m}.
\label{E8}
\end{equation}
$f_b(r)$ denotes the gravitational force of the disc \citep{YE06}
\begin{equation}
f_b(r)=-\left.\frac{dV_{j,k}}{dr}\right|_{r=r_m}=-2\int_{r'}\frac{\rho(r')r'}{r}\left[\frac{E(\zeta)}{r-r'}+\frac{F(\zeta)}{r+r'}\right]dr',
\label{E9}
\end{equation}
where $E(\zeta)$ is elliptic integral of the second kind. We calculate $f_b(r)$ at $r=r_m=0.99$. By expanding the elliptic integral in  equations (\ref{E6}) and (\ref{E9}) within the limit $k\leq r'\leq j$  and choosing an appropriate terms relative to $r$, we write a simplified form of $f_b(r)$ as \citep{YE06}:
\begin{equation}
f_b(r)=-\left.2ch\pi \frac{j-k}{jk}\frac{1}{r^2}-\frac{3}{8}ch\pi\frac{1}{r^3}\mbox{Log}\left[\frac{j}{k}\right]\right|_{r=r_m}.
\label{E10}
\end{equation}
We assume that the gravitational force is radially symmetric, so we have $\frac{x}{r}f_b(r)$ and $\frac{y}{r}f_b(r)$ as $x$ and $y$ components of the force $f_b(r)$ respectively\footnote{For detail derivation of equations (\ref{E8}, \ref{E9} and \ref{E10}), one is advised to check ref. \citep{YE06}}. The equations of motion of the infinitesimal mass are:
\begin{subequations}
\begin{equation}
\ddot{x}-2n\dot{y}=\Omega_x
\label{E11a}
\end{equation}
\begin{equation}
\ddot{y}+2n\dot{x}=\Omega_y,
\label{E11b}
\end{equation}
\end{subequations}
where
\begin{equation}
\Omega=\frac{n^2(x^2+y^2)}{2}+q_1(1-\mu)\left(\frac{1}{r_1}+\frac{A_1}{2r_1^3}-\frac{3A_2}{8r_1^5}\right)+q_2\mu \left(\frac{1}{r_2}+\frac{B_1}{2r_2^3}-\frac{3B_2}{8r_2^5}\right)-V_{j,k}.
\label{E12}
\end{equation}
It should be noted here that the suffixes $x$ and $y$ indicate the partial derivatives of $\Omega$ with respect to $x$ and $y$ respectively. This system admits the well-known Jacobi integral:
\begin{equation}
C=2\Omega-(\dot{x}^2+\dot{y}^2),
\label{E13}
\end{equation}
where $C$ is the Jacobi constant.
\section{Calculation of libration points}
libration points are very important in astronomy because they indicate places where particle can be trapped. In this section, we obtain the libration points which are the singularities of the manifold
\begin{equation}
C=2\Omega-(\dot{x}^2+\dot{y}^2).
\label{E14}
\end{equation}
To achieve this task, we equate all velocities and accelerations of the dynamical systems to zero (i.e., $\dot{x}=\dot{y}=\Omega_x=\Omega_y=0$). Thus,
\begin{subequations}
\begin{eqnarray}
\Omega_x&=&\frac{\partial \Omega}{\partial x}=n^2x-2ch\left(\frac{j-k}{jk}\right)\frac{\pi x}{r^3}-\frac{3ch\pi x}{8r^4}\mbox{Log}\left[\frac{j}{k}\right]\nonumber\\
        &&+\frac{q_1(\mu-1)(x+\mu)\left(-15A_2+12A_1r_1^2+8r_1^4\right)}{8r_1^7}\label{E15a}\\
				&&-\frac{\mu q_2(x+\mu-1)\left(-15B_2+12B_1r_2^2+8r_2^4\right)}{8r_2^7},\nonumber
\end{eqnarray}
\begin{eqnarray}
\Omega_y&=&\frac{\partial \Omega}{\partial y}=n^2y-2ch\left(\frac{j-k}{jk}\right)\frac{\pi y}{r^3}-\frac{3ch\pi y}{8r^4}\mbox{Log}\left[\frac{j}{k}\right]\nonumber\\
        &&+\frac{q_1(\mu-1)y\left(-15A_2+12A_1r_1^2+8r_1^4\right)}{8r_1^7}\label{E15b}\\
				&&-\frac{\mu q_2y\left(-15B_2+12B_1r_2^2+8r_2^4\right)}{8r_2^7}.\nonumber
\end{eqnarray}
\end{subequations}
Re-arranging equations (\ref{E15a}) and \ref{E15b}), we find
\begin{subequations}
\begin{eqnarray}	&&x\left(n^2-2ch\left(\frac{j-k}{jk}\right)\frac{\pi}{r^3}-\frac{3ch\pi}{8r^4}\mbox{Log}\left[\frac{j}{k}\right]+\frac{q_1(\mu-1)\left(-15A_2+12A_1r_1^2+8r_1^4\right)}{8r_1^7}\right.\nonumber\\
        &&\left.-\frac{\mu q_2\left(-15B_2+12B_1r_2^2+8r_2^4\right)}{8r_2^7}\right)+\left(\frac{q_1(\mu-1)\mu\left(-15A_2+12A_1r_1^2+8r_1^4\right)}{8r_1^7}\right.\label{E16a}\\
				&&\left.-\frac{\mu(\mu-1) q_2\left(-15B_2+12B_1r_2^2+8r_2^4\right)}{8r_2^7}\right)=0,\nonumber			
\end{eqnarray}
\begin{eqnarray} &&y\left(n^2-2ch\left(\frac{j-k}{jk}\right)\frac{\pi}{r^3}-\frac{3ch\pi}{8r^4}\mbox{Log}\left[\frac{j}{k}\right]+\frac{q_1(\mu-1)\left(-15A_2+12A_1r_1^2+8r_1^4\right)}{8r_1^7}\right.\label{E16b}\nonumber		\\
        &&\left.-\frac{\mu q_2\left(-15B_2+12B_1r_2^2+8r_2^4\right)}{8r_2^7}\right)=0.	
\end{eqnarray}
\end{subequations}
Now, we solve  equations (\ref{E16a}) and \ref{E16b}) for $y\neq 0$ to obtain the triangular libration points. From equation (\ref{E16b}), if $y\neq 0$, then it requires that  
\begin{eqnarray} &&n^2-2ch\left(\frac{j-k}{jk}\right)\frac{\pi}{r^3}-\frac{3ch\pi}{8r^4}\mbox{Log}\left[\frac{j}{k}\right]+\frac{q_1(\mu-1)\left(-15A_2+12A_1r_1^2+8r_1^4\right)}{8r_1^7}\nonumber\\
        &&-\frac{\mu q_2\left(-15B_2+12B_1r_2^2+8r_2^4\right)}{8r_2^7}=0.
				\label{E17}
\end{eqnarray}
By substituting equation (\ref{E17}) in (\ref{E16a}), we find
\begin{equation}
\frac{q_1\left(-15A_2+12A_1r_1^2+8r_1^4\right)}{8r_1^7}-\frac{q_2\left(-15B_2+12B_1r_2^2+8r_2^4\right)}{8r_2^7}=0.
\label{E18}
\end{equation}
Again, we re-write equation (\ref{E17}) in the form of
\begin{eqnarray} &&n^2-2ch\left(\frac{j-k}{jk}\right)\frac{\pi}{r^3}-\frac{3ch\pi}{8r^4}\mbox{Log}\left[\frac{j}{k}\right]-\frac{q_1\left(-15A_2+12A_1r_1^2+8r_1^4\right)}{8r_1^7}\nonumber\\
&&+\mu\left(\frac{q_1\left(-15A_2+12A_1r_1^2+8r_1^4\right)}{8r_1^7}-\frac{q_2\left(-15B_2+12B_1r_2^2+8r_2^4\right)}{8r_2^7}\right)=0.
\label{E19}
\end{eqnarray}
Substituting equation (\ref{E18}) into (\ref{E19}), we find
\begin{equation}
n^2-2ch\left(\frac{j-k}{jk}\right)\frac{\pi}{r^3}-\frac{3ch\pi}{8r^4}\mbox{Log}\left[\frac{j}{k}\right]-\frac{q_1\left(-15A_2+12A_1r_1^2+8r_1^4\right)}{8r_1^7}=0.
\label{E20}
\end{equation}
From equations (\ref{E18}) and (\ref{E20}), we obtain
\begin{equation}
n^2-2ch\left(\frac{j-k}{jk}\right)\frac{\pi}{r^3}-\frac{3ch\pi}{8r^4}\mbox{Log}\left[\frac{j}{k}\right]-\frac{q_2\left(-15B_2+12B_1r_2^2+8r_2^4\right)}{8r_2^7}=0.
\label{E21}
\end{equation}
By neglecting the effects of radiation and oblateness of the primaries, the above equations (\ref{E20}) and (\ref{E21}) reduce to classical case with the solutions $r_1 = r_2 = 1$. Hence, it is reasonable if we assume the solutions of equations (\ref{E20}) and (\ref{E21}) as $r_1=1+\epsilon_1$ and  $r_2=1+\epsilon_2$ respectively, where $\epsilon_1$, $\epsilon_2$ are very small quantities. Using equations (\ref{E8}), (\ref{E10}) and (\ref{E20}), we find
\begin{equation}
1+\frac{3}{2}\left(A_1+B_1-\frac{5}{4}(A_2+B_2)\right)-\left.f_b(r)\right|_{r=r_m}-\frac{q_1\left(-15A_2+12A_1r_1^2+8r_1^4\right)}{8r_1^7}=0.
\label{E22}
\end{equation}
Thus, since $\epsilon_i$ are very small quantities, hence, $(t)/r_1^7$, $(t)/r_1^5$ and $(t)/r_1^3$ can all be approximated as $t$ ($\approx t$), where $t$ could be any parameter in the numerator of the last expression of equation (\ref{E22}). Now, substituting $q_1=1-p_1$, in equation (\ref{E22})  and then restricting ourselves to the linear terms in $\epsilon_i$, $A_i$, $p_i$, and coupling terms in $q_1A_i$, we have
\begin{equation}
1+\frac{3}{2}\left(A_1(1-q_1)+B_1\right)-\frac{15}{8}\left(A_2(1-q_1)+B_2\right)-\left.f_b(r)\right|_{r=r_m}=\frac{1}{r_1^3}-p_1,
\label{E23}
\end{equation}
or
\begin{eqnarray}
r_1&=&\left(1+p_1+\frac{3}{2}\left(A_1(1-q_1)+B_1\right)-\frac{15}{8}\left(A_2(1-q_1)+B_2\right)-\left.f_b(r)\right|_{r=r_m}\right)^{-\frac{1}{3}}\nonumber\\
&=&1+\left(-\frac{p_1}{3}-\frac{1}{2}\left(A_1p_1+B_1\right)+\frac{5}{8}\left(A_2p_1+B_2\right)+\frac{1}{3}\left.f_b(r)\right|_{r=r_m}\right)\nonumber\\
&=&1+\epsilon_1.
\label{E24}
\end{eqnarray}
Using similar approach,
\begin{eqnarray}
r_2&=&1+\left(-\frac{p_2}{3}-\frac{1}{2}\left(B_1p_2+A_1\right)+\frac{5}{8}\left(B_2p_2+A_2\right)+\frac{1}{3}\left.f_b(r)\right|_{r=r_m}\right)\nonumber\\
   &=&1+\epsilon_2.
\label{E25}
\end{eqnarray}
Substituting equations (\ref{E22}) and (\ref{E23}) into equation (\ref{E4}) with restriction to the linear terms in $\epsilon_i$, we find
\begin{eqnarray}
x_0&=&\frac{1}{2}\left(r_1^2-r_2^2-2\mu+1\right)=\epsilon_1-\epsilon_2-\mu+\frac{1}{2},\nonumber\\
y_0&=&\left(r_1^2-(x+\mu)^2\right)^{1/2}=\left(r_1^2-\frac{1}{4}\left(1+2\epsilon_1-2\epsilon_2\right)^2\right)^{1/2}\nonumber\\
&=&\frac{\sqrt{3}}{2}\left(1+\frac{4}{3}\epsilon_1+\frac{4}{3}\epsilon_2\right)^{1/2}\approx\frac{\sqrt{3}}{2}\left(1+\frac{2}{3}\epsilon_1+\frac{2}{3}\epsilon_2\right).
\label{E26}
\end{eqnarray}
\begin{table}[!t]
{\scriptsize
\caption{Effect of the perturbations and PDP on the positions of the triangular libration points $L_{4,5}$ with $\mu=0.4$, $r_m=0.99$, $j = 1.5$, $k = 1$, $h = 0.0001$ and $c = 1910.83$} \vspace*{10pt}{
\begin{tabular}{ccccccccccc}\\[1.5ex]\hline\hline
\multicolumn{7}{c}{}&\multicolumn{1}{c}{PDP $=0$}&\multicolumn{2}{c}{}&\multicolumn{1}{c}{PDP $\neq0$}\\[1.5ex]
{}&{}&{}&{}&{}&{}&{}&{}&{}&{}&{}\\[-1.0ex]
Case&$p_1$&$p_2$&$A_1$&$B_1$&$A_2$&$B_2$	    &L$_{4,5}$      &&&        L$_{4,5}$\\[2.5ex]\hline\hline
1	&0		&0		&0		&0		&0		&0		&0.100000, $\pm$0.866025&&&0.100000,    $\pm$0.693135	\\[2ex]\hline
	&			&			&			&			&			&			&		&	\\[1ex]
2	&0		&0		&0.03	&0		&0.02	&0		&0.102500, $\pm$0.864582&&&0.102500, $\pm$0.691691	\\[1ex]
	&0		&0		&0.04	&0		&0.03	&0		&0.101250, $\pm$0.865304&&&0.101250, $\pm$0.692413	\\[1ex]
	&0		&0		&0.05	&0		&0.04	&0		&0.100000, $\pm$0.866025&&&0.100000, $\pm$0.693135	\\[2ex]\hline
	&			&			&			&			&			&			&												&&&													\\[1ex]
3	&0		&0		&0		&0.03	&0		&0.02	&0.108750, $\pm$0.871077&&&0.108750, $\pm$0.698187	\\[1ex]
	&0		&0		&0		&0.04	&0		&0.03	&0.110000, $\pm$0.871798&&&0.110000, $\pm$0.698908	\\[1ex]
	&0		&0		&0		&0.05	&0		&0.04	&0.111250, $\pm$0.872521&&&0.111250, $\pm$0.699630	\\[2ex]\hline
	&			&			&			&			&			&	    &												&&&													\\[1ex]
4	&0.10	&0		&0		&0		&0		&0		&0.066667, $\pm$0.846780&&&0.666667, $\pm$0.673890	\\[1ex]
	&0.15	&0		&0		&0		&0		&0	  &0.050000, $\pm$0.837157&&&0.500000, $\pm$0.664267	\\[1ex]
	&0.20	&0		&0		&0		&0		&0	  &0.033333, $\pm$0.827535&&&0.333333, $\pm$0.654645	\\[2ex]\hline
	&			&			&			&			&			&			&												&&&													\\[1ex]
5	&0		&0.15	&0		&0		&0		&0	  &0.150000, $\pm$0.837157&&&0.150000, $\pm$0.664267	\\[1ex]
	&0		&0.25	&0		&0		&0		&0	  &0.183333, $\pm$0.817912&&&0.183333, $\pm$0.645022	\\[1ex]
	&0		&0.35	&0		&0		&0		&0	  &0.216667, $\pm$0.798667&&&0.216667, $\pm$0.625777	\\[2ex]\hline
	&			&			&			&			&			&			&												&&&													\\[1ex]
6	&0.10	&0.15	&0.03	&0.02	&0.03	&0.02	&0.115417, $\pm$0.821954&&&0.115417, $\pm$0.649064	\\[1ex]
	&0.15	&0.25	&0.04	&0.03	&0.04	&0.03	&0.131896, $\pm$0.795071&&&0.131896, $\pm$0.622181	\\[1ex]
	&0.20	&0.35	&0.05	&0.04	&0.05	&0.04	&0.148250, $\pm$0.768405&&&0.148250, $\pm$0.595514	\\[2ex]
\hline\hline
\end{tabular}\label{Tab1}}
\vspace*{-1pt}}
\end{table}
Equation (\ref{E26}) has been used to compute the results in Table \ref{Tab1}. As it can be seen from the table, the coordinates of the triangular libration points $L_4$ and $L_5$  are affected by the oblateness and the radiation of the primaries when PDP is $0$ and $\neq$0. If from equation (\ref{E26}), we neglect coupling terms (CT) and PDP (i.e. setting $A_ip_i = B_ip_i = c = 0$, $i=1,2$) our results are in excellent agreement with the ones obtained in ref. \citep{YE1} for a special case $M_b=0$. Also, our results coincide with that of ref. \citep{YE16} for $M_b=0$ by neglecting  the effects of radiation of the primaries, PDP and oblateness of the less massive primary (i.e. setting $p_1 = p_2 = c = B_1 = B_2 = 0$). Furthermore, by considering only linear terms in small quantities in ref. \citep{YE17} and if we ignore the effects of oblateness of the more massive primary, PDP, the radiation and the oblateness coefficient $B_2$ of the less massive primary in the current study (i.e. setting $p_2 = A_1 = A_2 = B_2 = 0$), the libration points in the current results agree with those obtained  in ref. \citep{YE17} for a special case $M_b=0$

For case $1$ (as presented in Table \ref{Tab1}), in the absence of the oblateness, the radiation of the primaries and PDP, our results agree with that of Singh and Taura \citep{YE4}. In general our results are in excellent agreement with the ones obtained (presented as case $1-7$) by these authors if we set $A_2=B_2=c=0$ except for their cases $2$ and $3$ which were due to erroneous negative in the coefficient of $(A_1+A_2)$ in their equation (14). For case $2$,  we consider only the oblateness of the more massive primary (i.e. we set $p_1 = p_2 = B_1 = B_2 = 0$) and in the absence of PDP (i.e. $c = 0$). The triangular libration points $L_4$, $L_5$ come nearer to the primaries and the line joining the primaries. The same behavior is also experience in the presence of power profile. For case $3$, we considered only the oblateness of less massive primary (i.e. $q_1 = q_2 = 1, A_1 = A_2 = 0$). Consequently, the triangular libration points move away from the line joining the primaries.

Generally speaking, the presence of power-law profile does not affect the $x-$component of the libration points. On considering only the radiation of the less massive primary (i.e. setting $p_1 = A_1 = A_2 = B_1 = B_2 = 0$), the libration points tend towards the line joining the primaries. This can be seen in case $4$. Furthermore, on considering as case $5$, only the radiation of the more massive primary (i.e. setting $p_2 = A_1 = A_2 = B_1 = B_2 = 0$), the libration points move away from the line joining the primaries. As a final case, we explored the overall effect due to all the perturbations. We find that $L_{4,5}$, move towards the less massive primary. It should be noted that cases where power-law profile is considered exhibit similar behavior with cases where power-law profile is $0$.

It is worth mentioning that the comprehensive effects of the perturbations have a stabilizing proclivity. However, the oblateness up to $J_2$ of the primaries and the radiation of the primaries have tendency for instability, while coefficients up to $J_4$ of the primaries have stability predisposition.

\section{Linear Stability}
To study the stability of a libration points $(x_0,y_0)$, we apply infinitesimal displacement $\zeta$ and $\eta$ to the coordinates by using $\eta=y-y_0$ and $\zeta=x-x_0$ and then substitute them  into equations of motion in (\ref{E11a}) and  (\ref{E11b}) to obtain
\begin{equation}
\ddot{\zeta}-2n\dot{\eta}=\zeta\Omega_{xx}^0+\eta\Omega_{xy}^0\ \ \ \ \ \ \ddot{\eta}+2n\dot{\zeta}=\zeta\Omega_{yx}^0+\eta\Omega_{yy}^0,
\label{E27}
\end{equation}
where the superfix $`0'$ indicates that the partial derivatives have been computed at the triangular libration points by considering $(x_0,y_0)$. Now, let us assume a solution of the form $\zeta=C_1\exp^{\lambda t}$ and $\eta=C_2\exp^{\lambda t}$, where $C_1$ and $C_2$ are constants and $\lambda$ is a parameter. Substituting the assumed solutions into equation (\ref{E27}), we obtain the following non-trivial solutions for $C_1$ and $C_2$
\begin{equation}
\left|\begin{matrix}\lambda^2-\Omega_{xx}^0&&-2n\lambda-\Omega_{xy}^0\\ 2n\lambda-\Omega_{xy}^0&&\lambda^2-\Omega_{yy}^0 \end{matrix}\right|=0.
\label{E28}
\end{equation}
Solving the determinant by expansion, the characteristic equation corresponding to the variational equations (\ref{E27}) can be found as:
\begin{equation}
\lambda^4+(4n^2-\Omega_{xx}^0-\Omega_{yy}^0)\lambda^2+\left(\Omega_{xx}^0\Omega_{yy}^0-{\Omega_{xy}^0}^2\right)=0.
\label{E29}
\end{equation}
Now, we find the second order derivatives of the potential function w.r.t $x$ and $y$ and we write the expressions as follows:
\begin{subequations}
\begin{eqnarray}
\Omega_{xx}&=&n^2+\frac{6ch\pi x^2}{r^5}\frac{j-k}{jk}-\frac{2ch\pi}{r^3}\frac{j-k}{jk}+\left[\frac{3ch\pi x^2}{2r^6}-\frac{3ch\pi}{8r^4}\right]\text{Log}\left(\frac{j}{k}\right)\nonumber\\
&&+\frac{(\mu-1){q_1}\left(-48{A_1}{r_1}^4+60{A_1}{r_1}^2y^2+15{A_2}\left(6{r_1}^2-7y^2\right)-16{r_1}^6+24{r_1}^4y^2\right)}{8{r_1}^9}\nonumber\\
&&+\frac{\mu{q_2} \left(48{B_1}{r_2}^4-60{B_1}{r_2}^2 y^2-90{B_2}{r2}^2+105{B_2} y^2+16{r_2}^6-24{r_2}^4 y^2\right)}{8{r_2}^9},\label{E30a}\\
\Omega_{yy}&=&n^2+\left[\frac{3ch\pi y^2}{2r^6}-\frac{3ch\pi}{8r^4}\right]+\text{Log}\left(\frac{j}{k}\right)\frac{6ch\pi y^2}{r^5}\frac{j-k}{jk}-\frac{2ch\pi}{r^3}\frac{j-k}{jk}\nonumber\\
&&+q_1(\mu-1)\left[\frac{4{r_1}^2\left[3{A_1}\left({r_1}^2-5 y^2\right)+2 {r_1}^2\left({r_1}^2-3 y^2\right)\right]-15 {A_2} \left({r_1}^2-7 y^2\right)}{8 {r_1}^9}\right],\label{E30b}\\
&&+q_2\mu\left[\frac{15 {B_2} \left({r_2}^2-7 y^2\right)-4 {r_2}^2 \left[3{B_1}\left({r_2}^2-5y^2\right)+2 {r_2}^2 \left({r_2}^2-3 y^2\right)\right]}{8 {r_2}^9}\right],\nonumber\\
\Omega_{xy}&=&\frac{3}{2}xy\left[\frac{4ch\pi}{r^5}\frac{j-k}{jk}+\frac{ch\pi}{r^6}\text{Log}\left(\frac{j}{k}\right)\right]+3\mu q_2\left[\frac{\left(-35B_2+20B_1r_2^2+8r_2^4\right)(\mu+x-1)y}{8r_1^9}\right]\nonumber\\
&&-3(\mu-1)q_1\left[\frac{\left(-35A_2+20A_1r_1^2+8r_1^4\right)(\mu+x)y}{8r_1^9}\right].\label{E30c}
\end{eqnarray}
\end{subequations}
To get $\Omega^{0}_{xx}$, $\Omega^{0}_{yy}$  and $\Omega^{0}_{xy}$, we substitute the libration points given by equations (\ref{E26}) into equations (\ref{E30a}, \ref{E30b} and \ref{E30c}). Furthermore, we restrict our calculations to only the linear terms in $\epsilon_i$, $A_i$, $p_i$ and coupling terms in $q_iA_i$ and $q_iB_i$. Thus, we obtain
\begin{subequations}
\begin{eqnarray}
\Omega^{0}_{xx}&=&(\mu-1)(1-{p_1})\left[\left({\epsilon_1}+{\epsilon_2}+\frac{3}{4}\right) \left(\frac{15 {A_1}}{2} (1-7 {\epsilon_1})-\frac{105 {A_2}}{8}(1-9 {\epsilon_1})+3(1-5 {\epsilon_1})\right)\right.\nonumber\\
&&\left.-6{A_1} (1-5 {\epsilon_1})+\frac{1}{4} (45 {A_2}) (1-7 {\epsilon_1})-2 (1-3 {\epsilon_1})\right]-\frac{3\pi ch}{8 {r_m}^4} \mbox{Log}\left[\frac{j}{k}\right]-\frac{2\pi ch}{{ro}^3}\frac{j-k}{jk}\nonumber\\
&&+\mu(1-{p_2})\left[\left({\epsilon_1}+{\epsilon_2}+\frac{3}{4}\right) \left(-\frac{15 {B_1}}{2} (1-7 {\epsilon_2})+\frac{105 {B_2}}{8}(1-9 {\epsilon_2})-3 (1-5 {\epsilon_2})\right)\right.\nonumber\\
&&\left.+6 {B_1} (1-{\epsilon_2})-\frac{45 {B_2}}{4}(1-7 {\epsilon_2})+2 (1-3 {\epsilon_2})\right]+n^2\nonumber\\
&&+\left({\epsilon_1}-{\epsilon_2}+\mu^2-\mu+\frac{1}{4}\right) \left[\frac{3\pi ch}{2{r_m}^6} \mbox{Log}\left[\frac{j}{k}\right]+\frac{6\pi  ch}{{r_m}^5}\frac{j-k}{jk}\right],
\end{eqnarray}
\begin{eqnarray}
\Omega^{0}_{yy}&=&n^2+\left(\frac{3}{4}+\epsilon_1+\epsilon_2\right)\left(\frac{6ch\pi}{r_m^5}\frac{j-k}{jk}+\frac{3ch\pi}{2r_m^6}\mbox{Log}\left[\frac{j}{k}\right]\right)+\mu(1-p_2)\nonumber\\
&&\times\left[\left({\epsilon_1}+{\epsilon_2}+\frac{3}{4}\right) \left(-\frac{105}{8}B_2 (1-9\epsilon_2)+\frac{15}{2}B_1(1-7\epsilon_2)+3(1-5\epsilon_2)\right)\right.\nonumber\\
&&\left.+\frac{15}{8}B_2(1-7\epsilon_2)-\frac{3}{2}B_1(1-7\epsilon_2)-(1-3\epsilon_2)\right]-\frac{2ch\pi}{r_m^3}\frac{j-k}{jk}-\frac{3ch\pi}{8r_m^4}\mbox{Log}\left[\frac{j}{k}\right]\nonumber\\
&&+(\mu-1)(1-p_1)\left[\left({\epsilon_1}+{\epsilon_2}+\frac{3}{4}\right) \left(\frac{105}{8}A_2 (1-9\epsilon_1)-\frac{15}{2}A_1(1-7\epsilon_1)-3(1-5\epsilon_1)\right)\right.\nonumber\\
&&\left.-\frac{15}{8}A_2(1-7\epsilon_1)+\frac{3}{2}A_1(1-5\epsilon_1)+(1-3\epsilon_1)\right],
\end{eqnarray}
\begin{eqnarray}
\Omega^{0}_{xy}&=&\frac{3}{2}\frac{\sqrt{3}}{2}\left(1+\frac{2}{3}\epsilon_1+\frac{2}{3}\epsilon_2\right)\left(\epsilon_1-\epsilon_2-\mu+\frac{1}{2}\right)\left(\frac{4ch\pi}{r_m^5}\frac{j-k}{jk}+\frac{ch\pi}{r_m^6}\mbox{Log}\left[\frac{j}{k}\right]\right)\nonumber\\
&&+(\mu-1)(1-{p_1})\left(1+\frac{2}{3}\epsilon_1+\frac{2}{3}\epsilon_2\right)\left(\epsilon_1-\epsilon_2-\mu+\frac{1}{2}\right)\left(\frac{105}{8}A_2(1-9\epsilon_1)\right.\nonumber\\
&&\left.-\frac{15}{2}A_1(1-7\epsilon_1)-3(1-5\epsilon_1)\right)+\mu(1-{p_2})\left(1+\frac{2}{3}\epsilon_1+\frac{2}{3}\epsilon_2\right)\left(\epsilon_1-\epsilon_2-\frac{1}{2}\right)
\nonumber\\
&&\times\left(-\frac{105}{8}B_2(1-9\epsilon_2)+\frac{15}{2}B_1(1-7\epsilon_2)+3(1-5\epsilon_2)\right).
\end{eqnarray}
\end{subequations}
More explicitly, we can re-write the above expressions for $\Omega^{0}_{xx}$, $\Omega^{0}_{yy}$  and $\Omega^{0}_{xy}$ as
\begin{subequations}
\begin{eqnarray}
\Omega^{0}_{xx}&=&\frac{3}{4}+\frac{ch\pi}{2}\frac{j-k}{jk}t_2+\alpha_1+\mu\left[\beta_1-\frac{3ch\pi}{r_m^5}\left(2\frac{j-k}{jk}+\frac{1}{2r_m} \mbox{Log}\left[\frac{j}{k}\right]\right)\right]\\
&&+\frac{3}{8}t_1 ch\pi \mbox{Log}\left[\frac{j}{k}\right],\nonumber\\
\Omega^{0}_{xx}&=&\frac{9}{4}+\frac{ch\pi}{2}\frac{j-k}{jk}v_1+\frac{3}{8}v_2ch\pi\mbox{Log}\left[\frac{j}{k}\right]+\alpha_2+\mu\beta_2,\\
\Omega^{0}_{xy}&=&\mu\left[\beta_3-\frac{3\sqrt{3}}{2}-\sqrt{3}ch\pi\frac{j-k}{jk}\left(\frac{3}{r_m^5}+\frac{11}{3r_m^2}\right)+\left(\frac{3}{4r_m^6}-\frac{11}{16r_m^3}\right)\sqrt{3}ch\pi\frac{j-k}{jk}\right]\\
&&+\frac{3\sqrt{3}}{4}+w_1\sqrt{3}ch\pi \mbox{Log}\left[\frac{j}{k}\right]+\frac{ch\pi}{2}\frac{j-k}{jk}w_2,\nonumber
\end{eqnarray}
\end{subequations}
where we have introduced the following notations for mathematical simplicity
\begin{eqnarray}
t_1&=&\frac{1}{r_m^6}-\frac{1}{r_m^4}+\frac{9}{4r_m^3},\ \ \ \ \ v_1=\frac{9}{r_m^5}-\frac{4}{r_m^3}+\frac{11}{r_m^2}, \ \ \ \ \ \ \ w_1=\frac{3}{8r_m^6}+\frac{11}{32r_m^3},\nonumber\\
t_2&=&\frac{3}{r_m^5}-\frac{4}{r_m^3}+\frac{9}{r_m^2},\ \ \ \ \ \ v_2=\frac{3}{r_m^6}-\frac{1}{r_m^4}+\frac{11}{4r_m^3},\ \ \ \ \ \ \ w_2=\frac{3\sqrt{3}}{r_m^5}+\frac{11}{\sqrt{3}r_m^2},\nonumber\\
\end{eqnarray}
and $\beta_i$, $\alpha_i$ ($i=1,2,3$) represent:
\begin{eqnarray}
\beta_1&=&A_1\left[-3+\frac{25ch\pi}{4r_m^2}\frac{j-k}{jk}+\frac{75ch\pi}{64r_m^3}\mbox{Log}\left[\frac{j}{k}\right]+\frac{41}{8}p_1+\frac{29}{8}p_2\right]+B_1\left[3-\frac{89ch\pi}{4r_m^2}\frac{j-k}{jk}\right.\nonumber\\
&&\left.-\frac{267ch\pi}{64r_m^3}\mbox{Log}\left[\frac{j}{k}\right]-\frac{29}{8}p_1-\frac{105}{8}p_2\right]+A_2\left[\frac{75}{16}-\frac{25ch\pi}{16r_m^2}\frac{j-k}{jk}-\frac{75ch\pi}{256r_m^3}\mbox{Log}\left[\frac{j}{k}\right]\right.\nonumber\\
&&\left.+\frac{41}{8}p_1+\frac{29}{8}p_2\right]+B_2\left[-\frac{75}{16}+\frac{25ch\pi}{16r_m^2}\frac{j-k}{jk}+\frac{75ch\pi}{256r_m^3}\mbox{Log}\left[\frac{j}{k}\right]+\frac{105}{32}p_1+\frac{175}{32}p_2\right]\nonumber\\
&&+p_1\left[\frac{3}{2}+\frac{ch\pi}{2r_m^2}\frac{j-k}{jk}+\frac{3ch\pi}{32r_m^3}\mbox{Log}\left[\frac{j}{k}\right]\right]-p_2\left[\frac{3}{2}+\frac{ch\pi}{2r_m^2}\frac{j-k}{jk}+\frac{3ch\pi}{32r_m^3}\mbox{Log}\left[\frac{j}{k}\right]\right],\nonumber
\end{eqnarray}
\begin{eqnarray}
\alpha_1&=&A_1\left[\frac{27}{8}+\left(\frac{3}{4r_m^5}+\frac{35}{4r_m^2}\right)ch\pi\frac{j-k}{jk}+\left(\frac{3}{4r_m^6}+\frac{105}{64r_m^3}\right)ch\pi\mbox{Log}\left[\frac{j}{k}\right]-\frac{9}{8}p_1+\frac{5}{2}p_2\right]\nonumber\\
&&B_1\left[\frac{3}{8}-\left(\frac{1}{r_m^5}-\frac{5}{r_m^2}\right)3ch\pi\frac{j-k}{jk}-\left(\frac{3}{4r_m^6}-\frac{45}{16r_m^3}\right)ch\pi\mbox{Log}\left[\frac{j}{k}\right]+\frac{49}{8}p_1+4p_2\right]\nonumber\\
&&+A_2\left[-\frac{165}{32}-\left(\frac{15}{4r_m^5}+\frac{275}{16r_m^2}\right)ch\pi\frac{j-k}{jk}-\left(\frac{15}{16r_m^6}+\frac{825}{256r_m^3}\right)ch\pi\mbox{Log}\left[\frac{j}{k}\right]+\frac{15}{16}p_1-\frac{35}{8}p_2\right]\nonumber\\
&&+B_2\left[-\frac{15}{32}+\left(\frac{15}{4r_m^5}-\frac{75}{4r_m^2}\right)ch\pi\frac{j-k}{jk}+\left(\frac{15}{16r_m^6}+\frac{225}{64r_m^3}\right)ch\pi\mbox{Log}\left[\frac{j}{k}\right]-\frac{245}{32}p_1-5p_2\right]\nonumber\\
&&+p_1\left[-\frac{1}{2}-\left(\frac{2}{r_m^5}-\frac{19}{2r_m^2}\right)ch\pi\frac{j-k}{jk}-\left(\frac{1}{2r_m^6}-\frac{57}{32r_m^3}\right)ch\pi\mbox{Log}\left[\frac{j}{k}\right]\right]\nonumber\\
&&+p_2\left[1+\left(\frac{2}{r_m^5}+\frac{10}{3r_m^2}\right)ch\pi\frac{j-k}{jk}+\left(\frac{1}{2r_m^6}+\frac{5}{8r_m^3}\right)ch\pi\mbox{Log}\left[\frac{j}{k}\right]\right],\nonumber
\end{eqnarray}
\begin{eqnarray}
\beta_2&=&A_1\left[-\frac{85ch\pi}{4r_m^2}\frac{j-k}{jk}-\frac{255ch\pi}{64r_m^3}\mbox{Log}\left[\frac{j}{k}\right]-\frac{77}{8}p_1-\frac{41}{8}p_2\right]+B_1\left[\frac{77ch\pi}{4r_m^2}\frac{j-k}{jk}\right.\nonumber\\
&&\left.+\frac{231ch\pi}{64r_m^3}\mbox{Log}\left[\frac{j}{k}\right]+\frac{41}{8}p_1+\frac{69}{8}p_2\right]+A_2\left[\frac{45}{16}+\frac{725ch\pi}{16r_m^2}\frac{j-k}{jk}+\frac{2175ch\pi}{256r_m^3}\mbox{Log}\left[\frac{j}{k}\right]\right.\nonumber\\
&&\left.+\frac{635}{32}p_1+\frac{165}{32}p_2\right]+B_2\left[-\frac{45}{16}-\frac{725ch\pi}{16r_m^2}\frac{j-k}{jk}-\frac{2175ch\pi}{256r_m^3}\mbox{Log}\left[\frac{j}{k}\right]-\frac{165}{32}p_1-\frac{635}{32}p_2\right]\nonumber\\
&&+p_1\left[-\frac{3}{2}+\frac{3ch\pi}{2r_m^2}\frac{j-k}{jk}+\frac{9ch\pi}{32r_m^3}\mbox{Log}\left[\frac{j}{k}\right]\right]+p_2\left[\frac{3}{2}-\frac{3ch\pi}{2r_m^2}\frac{j-k}{jk}-\frac{9ch\pi}{32r_m^3}\mbox{Log}\left[\frac{j}{k}\right]\right],\nonumber
\end{eqnarray}
\begin{eqnarray}
\alpha_2&=&A_1\left[\frac{33}{8}-\left(\frac{3}{r_m^5}-\frac{25}{4r_m^2}\right)ch\pi\frac{j-k}{jk}-\left(\frac{3}{4r_m^6}-\frac{75}{64r_m^3}\right)ch\pi\mbox{Log}\left[\frac{j}{k}\right]+\frac{45}{8}p_1-\frac{5}{2}p_2\right]\nonumber\\
&&B_1\left[\frac{33}{8}-\left(\frac{1}{r_m^5}+\frac{5}{r_m^2}\right)3ch\pi\frac{j-k}{jk}-\left(\frac{1}{4r_m^6}+\frac{15}{16r_m^3}\right)3ch\pi\mbox{Log}\left[\frac{j}{k}\right]-\frac{61}{8}p_1-4p_2\right]\nonumber\\
&&+A_2\left[-\frac{255}{32}+\left(\frac{15}{4r_m^5}-\frac{425}{16r_m^2}\right)ch\pi\frac{j-k}{jk}+\left(\frac{15}{16r_m^6}-\frac{1275}{256r_m^3}\right)ch\pi\mbox{Log}\left[\frac{j}{k}\right]-\frac{475}{32}p_1+\frac{35}{8}p_2\right]\nonumber\\
&&+B_2\left[-\frac{165}{32}+\left(\frac{1}{4r_m^5}+\frac{5}{4r_m^2}\right)15ch\pi\frac{j-k}{jk}+\left(\frac{15}{16r_m^6}+\frac{225}{64r_m^3}\right)ch\pi\mbox{Log}\left[\frac{j}{k}\right]+\frac{105}{32}p_1+5p_2\right]\nonumber\\
&&+p_1\left[\frac{1}{2}-\left(\frac{2}{r_m^5}+\frac{23}{2r_m^2}\right)ch\pi\frac{j-k}{jk}-\left(\frac{1}{2r_m^6}+\frac{69}{32r_m^3}\right)ch\pi\mbox{Log}\left[\frac{j}{k}\right]\right]\nonumber\\
&&+p_2\left[-1-\left(\frac{2}{r_m^5}+\frac{10}{3r_m^2}\right)ch\pi\frac{j-k}{jk}-\left(\frac{1}{2r_m^6}+\frac{5}{8r_m^3}\right)ch\pi\mbox{Log}\left[\frac{j}{k}\right]\right],\nonumber
\end{eqnarray}
{\footnotesize
\begin{eqnarray}
\beta_3&=&A_1\left[-\frac{13\sqrt{3}}{4}+\left(\frac{\sqrt{3}}{r_m^5}-\frac{65}{4\sqrt{3}r_m^2}\right)ch\pi\frac{j-k}{jk}+\left(\frac{1}{4r_m^6}-\frac{125}{64r_m^3}\right)\sqrt{3}ch\pi\mbox{Log}\left[\frac{j}{k}\right]-\frac{27\sqrt{3}}{8}p_1-\frac{5}{2\sqrt{3}}p_2\right]\nonumber\\
&&B_1\left[-\frac{13\sqrt{3}}{4}+\left(\frac{\sqrt{3}}{r_m^5}-\frac{110}{4\sqrt{3}r_m^2}\right)ch\pi\frac{j-k}{jk}+\left(\frac{1}{4r_m^6}-\frac{125}{64r_m^3}\right)\sqrt{3}ch\pi\mbox{Log}\left[\frac{j}{k}\right]-\frac{5}{2\sqrt{3}}p_1-\frac{27\sqrt{3}}{8}p_2\right]\nonumber\\
&&+A_2\left[5\sqrt{3}-\left(\frac{5}{4r_m^5}-\frac{335}{16r_m^2}\right)\sqrt{3}ch\pi\frac{j-k}{jk}-\left(\frac{5}{16r_m^6}-\frac{1005}{256r_m^3}\right)\sqrt{3}ch\pi\mbox{Log}\left[\frac{j}{k}\right]+\frac{865}{32\sqrt{3}}p_1+\frac{35}{8\sqrt{3}}p_2\right]\nonumber\\
&&+B_2\left[5\sqrt{3}-\left(\frac{5}{4r_m^5}-\frac{335}{16r_m^2}\right)\sqrt{3}ch\pi\frac{j-k}{jk}-\left(\frac{5}{16r_m^6}-\frac{1005}{256r_m^3}\right)\sqrt{3}ch\pi\mbox{Log}\left[\frac{j}{k}\right]+\frac{35}{8\sqrt{3}}p_1+\frac{865}{32\sqrt{3}}p_2\right]\nonumber\\
&&-\frac{1}{2\sqrt{3}}(p_1+p_2),\nonumber
\end{eqnarray}
\begin{eqnarray}
\alpha_3&=&A_1\left[-\frac{19\sqrt{3}}{8}-\left(\frac{3}{r_m^5}-\frac{97}{4\sqrt{3}r_m^2}\right)ch\pi\frac{j-k}{jk}-\left(\frac{3}{4r_m^6}-\frac{97\sqrt{3}}{64r_m^3}\right)ch\pi\mbox{Log}\left[\frac{j}{k}\right]+\frac{49}{8\sqrt{3}}p_1+\frac{\sqrt{3}}{2}p_2\right]\nonumber\\
&&B_1\left[\frac{7\sqrt{3}}{8}-\left(\frac{3}{r_m^5}+\frac{13}{\sqrt{3}r_m^2}\right)ch\pi\frac{j-k}{jk}-\left(\frac{3}{4r_m^6}+\frac{13\sqrt{3}}{16r_m^3}\right)ch\pi\mbox{Log}\left[\frac{j}{k}\right]-\frac{31\sqrt{3}}{8}p_1+\frac{4}{\sqrt{3}}p_2\right]\nonumber\\
&&+A_2\left[-\frac{255}{32}+\left(\frac{15}{4r_m^5}-\frac{425}{16r_m^2}\right)ch\pi\frac{j-k}{jk}+\left(\frac{15}{16r_m^6}-\frac{1275}{256r_m^3}\right)ch\pi\mbox{Log}\left[\frac{j}{k}\right]-\frac{165\sqrt{3}}{32}p_1-\frac{25}{8\sqrt{3}}p_2\right]\nonumber\\
&&+B_2\left[-\frac{35\sqrt{3}}{32}+\left(\frac{15}{4r_m^5}+\frac{65}{4\sqrt{3}r_m^2}\right)ch\pi\frac{j-k}{jk}+\left(\frac{15}{16r_m^6}+\frac{65\sqrt{3}}{64r_m^3}\right)ch\pi\mbox{Log}\left[\frac{j}{k}\right]+\frac{155\sqrt{3}}{32}p_1-\frac{5}{\sqrt{3}}p_2\right]\nonumber\\
&&+p_1\left[-\frac{1}{2\sqrt{3}}-\left(\frac{4}{r_m^5}+\frac{85}{6r_m^2}\right)ch\pi\frac{j-k}{\sqrt{3}jk}-\left(\frac{1}{r_m^6}+\frac{85}{32r_m^3}\right)\frac{ch\pi}{\sqrt{3}}\mbox{Log}\left[\frac{j}{k}\right]\right]\nonumber\\
&&+p_2\left[\frac{1}{\sqrt{3}}+\left(\frac{1}{r_m^5}+\frac{1}{r_m^2}\right)2ch\pi\frac{j-k}{\sqrt{3}jk}+\left(\frac{1}{2\sqrt{3}r_m^6}+\frac{\sqrt{3}}{8r_m^3}\right)ch\pi\mbox{Log}\left[\frac{j}{k}\right]\right].
\end{eqnarray}}
Solving the characteristic equation (\ref{E29}) for $\lambda$, we obtain 
\begin{eqnarray}
\lambda_1=-\lambda_2=\sqrt{\frac{-\left(\Omega_{xx}^0+\Omega_{yy}^0-4n^2\right)+\sqrt{(\Omega_{xx}^0+\Omega_{yy}^0-4n^2)^2-4\left(\Omega_{xx}^0\Omega_{yy}^0-{\Omega_{xy}^0}^2\right)}}{2}}=\sqrt{\frac{-b+\sqrt{\delta}}{2}},\nonumber\\
\lambda_3=-\lambda_4=\sqrt{\frac{-\left(\Omega_{xx}^0+\Omega_{yy}^0-4n^2\right)-\sqrt{(\Omega_{xx}^0+\Omega_{yy}^0-4n^2)^2-4\left(\Omega_{xx}^0\Omega_{yy}^0-{\Omega_{xy}^0}^2\right)}}{2}}=\sqrt{\frac{-b-\sqrt{\delta}}{2}},\nonumber\\
\end{eqnarray}
with
\begin{eqnarray}
b&=&\Omega_{xx}^0+\Omega_{yy}^0-4n^2=-3+4n^2-(t_2+v_1)\frac{ch\pi}{2}\frac{j-k}{jk}-(\alpha_1+\alpha_2)-\frac{3}{8}(t_1+v_2)ch\pi\mbox{Log}\left[\frac{j}{k}\right]\nonumber\\
&&-\mu\left[\beta_1+\beta_2-\frac{3ch\pi}{r_m^5}\left(2\frac{j-k}{jk}+\frac{1}{2r_m}\mbox{Log}\left[\frac{j}{k}\right]\right)\right],\nonumber\\
\end{eqnarray}
where the discriminant $\delta$ is given by
\begin{eqnarray}
\delta&=&\left(\Omega_{xx}^0+\Omega_{yy}^0-4n^2\right)^2-4\left(\Omega_{xx}^0\Omega_{yy}^0-{\Omega_{xy}^0}^2\right)=-1+(1-8f_b(r_m))(-6+8(2f_b(r_m)+n^2))\nonumber\\
&&-11\alpha_1-5\alpha_2+6\sqrt{3}\alpha_3+\left\{t_2\left[-6-16n^2+4(\alpha_1-\alpha_2)\right]+v_1\left[6-16n^2-4(\alpha_1-\alpha_2)\right]\right.\nonumber\\
&&\left.+4w_2\left(3\sqrt{3}+4(\alpha_3)\right)\right\}\frac{ch\pi}{4}\frac{j-k}{jk}+\left\{3t_1\left[-6-16n^2+4(\alpha_1-\alpha_2)\right]+3v_2\left[6-16n^2\right.\right.\nonumber\\
&&\left.\left.-4(\alpha_1-\alpha_2)\right]+32w_1\left(9+4\sqrt{3}(\alpha_3)\right)\right\}\frac{ch\pi}{16}\mbox{Log}\left[\frac{j}{k}\right]+\mu^2\left\{3\left(9-4\sqrt{3}\beta_3\right)+\frac{2ch\pi}{3}\frac{j-k}{jk}\nonumber\right.\\
&&\left.\times\left(\frac{22}{r_m^2}\left(9-2\sqrt{3}\beta_3\right)+\frac{18}{r_m^5}\left(9-2\sqrt{3}\beta_3-\beta_1+\beta_2\right)\right)+\left[\frac{11}{r_m^3}\left(9-2\sqrt{3}\beta_3\right)\right.\right.\nonumber\\
&&\left.\left.-\frac{12}{r_m^6}\left(9-2\sqrt{3}\beta_3+\beta_1-\beta_2\right)\right]\frac{ch\pi}{4}\mbox{Log}\left[\frac{j}{k}\right]\right\}+\mu\left\{-27-12\sqrt{3}\alpha_3-11\beta_1-5\beta_2+6\sqrt{3}\beta_3\right.\nonumber\\
&&\left.+\frac{ch\pi}{3}\frac{j-k}{jk}\left[-\frac{22}{r_m^2}\left(9+4\sqrt{3}(\alpha_3)\right)\right.\right.+\left.\left.\frac{9}{r_m^5}\left(16n^2-4(3+\alpha_1-\alpha_2+2\sqrt{3}\alpha_3)\right)\right.\right.\nonumber\\
&&+\left.\left.3\left(-6\sqrt{3}w_2+(t_2-v_1)(\beta_1-\beta_2)+4w_2\beta_3\right)\right]+\left[-\frac{11}{r_m^3}\left(9+4\sqrt{3}(\alpha_3)\right)\right.\right.\nonumber\\
&&+\left.\left.\frac{6}{r_m^6}\left(16n^2+4(6-\alpha_1+\alpha_2+2\sqrt{3}\alpha_3)\right)+6(t_1-v_2)(\beta_1-\beta_2)+32w_1(-9+2\sqrt{3}\beta_3)\right]\right.\nonumber\\
&&\left.\times\frac{ch\pi}{8}\frac{j-k}{jk}\mbox{Log}\left[\frac{j}{k}\right]\right\}.\label{E37}
\end{eqnarray}
Now, using equation (\ref{E37}), we find
{\begin{eqnarray}
\left.\delta_\mu\right|_{\mu=0}&=&-27-12\sqrt{3}\alpha_3-11\beta_1-5\beta_2+6\sqrt{3}\beta_3+\frac{ch\pi}{3}\frac{j-k}{jk}\left[-\frac{22}{r_m^2}\left(9+4\sqrt{3}(\alpha_3)\right)\right.\nonumber\\
&&\left.+\frac{9}{r_m^5}\left(16n^2-4(3+\alpha_1-\alpha_2+2\sqrt{3}\alpha_3)\right)+3\left(-6\sqrt{3}w_2+(t_2-v_1)(\beta_1-\beta_2)+4w_2\beta_3\right)\right]\nonumber\\
&&+\frac{ch\pi}{8}\frac{j-k}{jk}\left[-\frac{11}{r_m^3}\left(9+4\sqrt{3}(\alpha_3)\right)+\frac{6}{r_m^6}\left(16n^2+4(6-\alpha_1+\alpha_2+2\sqrt{3}\alpha_3)\right)\right.\nonumber\\
&&\left.+6(t_1-v_2)(\beta_1-\beta_2)+32w_1(-9+2\sqrt{3}\beta_3)\right]\mbox{Log}\left[\frac{j}{k}\right]\approx-27<0.\label{E38}
\end{eqnarray}
\begin{eqnarray}
\left.\delta_\mu\right|_{\mu=\frac{1}{2}}&=&-12\sqrt{3}\alpha_3-11\beta_1-5\beta_2-6\sqrt{3}\beta_3+\frac{ch\pi}{3}\frac{j-k}{jk}\left[\frac{22}{r_m^2}\left(9-4\sqrt{3}(\alpha_3+\beta_3)\right)\right.\nonumber\\
&&+\left.\frac{9}{r_m^5}\left(16n^2+4(6-\alpha_1+\alpha_2-2\sqrt{3}\alpha_3+\beta_1-\beta_2+2\sqrt{3}\beta_3)\right)+3\left(w_2\left(-6\sqrt{3}+4\beta_3\right)\right.\right.\nonumber\\
&&+\left.\left.(t_2-v_1)(\beta_1-\beta_2)\right)\right]+\frac{ch\pi}{8}\frac{j-k}{jk}\left[\frac{11}{r_m^3}\left(9-4\sqrt{3}(\alpha_3+\beta_3)\right)+6(t_1-v_2)(\beta_1-\beta_2)\right.\nonumber\\
&&+\left.\frac{6}{r_m^6}\left(16n^2-4(3+\alpha_1-\alpha_2-2\sqrt{3}\alpha_3)\right)-32w_1(9-2\sqrt{3}\beta_3)\right]\mbox{Log}\left[\frac{j}{k}\right]\approx0.\label{E39}
\end{eqnarray}}
It can be deduced from (\ref{E37}) that $\left.\delta\right|_{\mu=0} >0$ and $\left.\delta\right|_{\mu=\frac{1}{2}}<0$ (i.e., the values of $\delta$ at the point $\mu=0$ and $\mu=\frac{1}{2}$ are of opposite signs). Also from equations (\ref{E38}) and (\ref{E39}), $\delta_\mu<0$ within the interval of $\left(0,\frac{1}{2}\right)$. Since $\delta$ is strictly decreasing function of $\mu$ in the open interval of $\left(0,\frac{1}{2}\right)$, thus, there is only one value of $\mu$ in the interval for which $\delta$ vanishes and it is denoted as $\mu_c$ (the critical mass parameter). Moreover, there exist three possible cases.  When $0<\mu<\mu_c$ ($\delta>0$), the values of $\lambda_{1,2,3,4}$  which are the solutions of the characteristic equation are pure imaginary numbers indicating that, triangular equilibrium points are linearly stable. Secondly, for $\mu=\mu_c$ $(\delta = 0)$, then $\lambda_{1,3}=\lambda_{2,4}$ (i.e., double roots). The solutions have a secular terms which gives the instability of the point. Finally, for $\mu_c\leq\mu\leq\frac{1}{2}$ ($\delta<0$), $\lambda_{1,2,3,4}$ have imaginary roots and two of the roots have positive real parts. Consequently, the triangular points are unstable. In the next section, we proceed to calculating $\mu_c$.
\section{Determination of $\mu_c$}
\begin{figure}[!t]
\includegraphics[height=115mm,width=180mm]{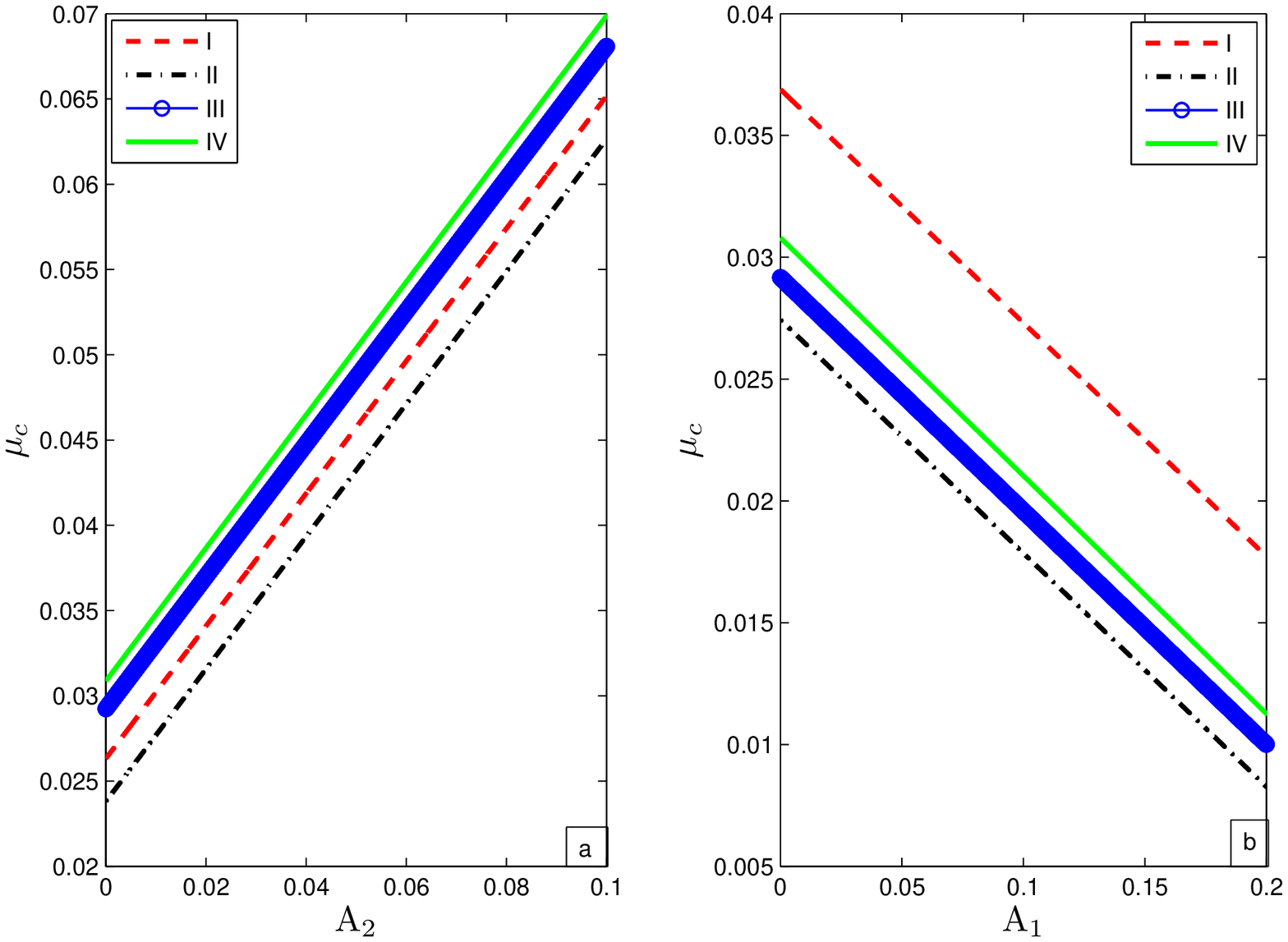}
\caption{{\protect\footnotesize (Color online) (a) Critical mass ratio as a function of $A_2$ with (I) $A_1 = 0.03$, $B_1 = 0.02$, $B_2 = 0.01$, $q_1 = 0.99$ and  $q_2=0.98$. (II) $A_1 = 0.03$, $B_1 = 0.02$, $B_2 = 0$, $q_1 = 0.99$ and  $q_2=0.98$. (III) $A_1 = B_1 = B_2 = 0$, $q_1 = 0.99$ and  $q_2=0.98$. (IV) $A_1 = B_1 = B_2 = p_1 = p_2=0$. (b) Critical mass ratio as a function of $A_1$ with (I)$B_1 = 0.02$ $A_2 = 0.02$, $B_2 = 0.01$, $q_1 = 0.99$ and  $q_2=0.98$. (II) $B_1 = 0.02$ $A_2 = B_2 = 0$, $q_1 = 0.99$ and  $q_2=0.98$. (III) $B_1 = A_2 = B_2 = 0$, $q_1 = 0.99$ and  $q_2=0.98$. (IV) $B_1 = A_2 = B_2 = p_1 = p_2 = 0$.}}
\label{fig2}
\end{figure}

A critical mass define a range for which the triangular libration points are stable or unstable. It could be a lower or upper limit for this range of values. To calculate the critical mass, we re-write the discriminant $\delta$ in the form
\begin{equation}
\delta=27(\mathcal{C}_{m_1}+1)\mu^2-27(\mathcal{C}_{m_2}+1)\mu+(\mathcal{C}_{m_3}+1),
\label{E40}
\end{equation}
where
\begin{eqnarray}
\mathcal{C}_{m_1}&=&\frac{1}{27}\left\{3\left(-4\sqrt{3}\beta_3\right)+\frac{2ch\pi}{3}\frac{j-k}{jk}\left(\frac{22}{r_m^2}\left(9-2\sqrt{3}\beta_3\right)+\frac{18}{r_m^5}\left(9-2\sqrt{3}\beta_3-\beta_1+\beta_2\right)\right)\right.\nonumber\\
&&\left.+\left[\frac{11}{r_m^3}\left(9-2\sqrt{3}\beta_3\right)-\frac{12}{r_m^6}\left(9-2\sqrt{3}\beta_3+\beta_1-\beta_2\right)\right]\frac{ch\pi}{4}\mbox{Log}\left[\frac{j}{k}\right]\right\},
\end{eqnarray}
\begin{eqnarray}
\mathcal{C}_{m_2}&=&-\frac{1}{27}\left\{-12\sqrt{3}\alpha_3-11\beta_1-5\beta_2+6\sqrt{3}\beta_3+\frac{ch\pi}{3}\frac{j-k}{jk}\left[-\frac{22}{r_m^2}\left(9+4\sqrt{3}(\alpha_3)\right)\right.\right.\nonumber\\
&&+\left.\left.\frac{9}{r_m^5}\left(16n^2-4(3+\alpha_1-\alpha_2+2\sqrt{3}\alpha_3)\right)+3\left(-6\sqrt{3}w_2+(t_2-v_1)(\beta_1-\beta_2)+4w_2\beta_3\right)\right]\right.\nonumber\\
&&\left.+\frac{ch\pi}{8}\frac{j-k}{jk}\left[-\frac{11}{r_m^3}\left(9+4\sqrt{3}(\alpha_3)\right)+\frac{6}{r_m^6}\left(16n^2+4(6-\alpha_1+\alpha_2+2\sqrt{3}\alpha_3)\right)\right.\right.\nonumber\\
&&+\left.\left.+6(t_1-v_2)(\beta_1-\beta_2)+32w_1(-9+2\sqrt{3}\beta_3)\right]\mbox{Log}\left[\frac{j}{k}\right]\right\}\nonumber\\
\end{eqnarray}
and
\begin{eqnarray}
\mathcal{C}_{m_3}&=&-11\alpha_1-5\alpha_2+6\sqrt{3}\alpha_3+\left\{t_2\left[-6-16n^2+4(\alpha_1-\alpha_2)\right]+v_1\left[6-16n^2\right.\right.\nonumber\\
&&\left.\left.-4(\alpha_1-\alpha_2)\right]+4w_2\left(3\sqrt{3}+4(\alpha_3)\right)\right\}\frac{ch\pi}{4}\frac{j-k}{jk}+\left\{3t_1\left[-6-16n^2+4(\alpha_1-\alpha_2)\right]\right.\nonumber\\
&&\left.+3v_2\left[6-16n^2-4(\alpha_1-\alpha_2)\right]+32w_1\left(9+4\sqrt{3}\alpha_3\right)\right\}\frac{ch\pi}{16}\mbox{Log}\left[\frac{j}{k}\right]\nonumber\\
&&-2+(1-8f_b(r_m))(-6+8(2f_b(r_m)+n^2)).
\label{E41}
\end{eqnarray}
It is very difficult to obtain an analytical solution of equation (\ref{E40}). To overcome this difficulty, Abouelmagd \citep{YE18} developed an algorithm to perform the calculation in a simplified form. Therefore, following ref. \citep{YE18}, we calculate $\mu_c$ as follows:
\begin{eqnarray}
\mu_c&=&\frac{27(1+\mathcal{C}_{m_2})-\sqrt{729(1+\mathcal{C}_{m_2})^2-108(1+\mathcal{C}_{m_1})(1+\mathcal{C}_{m_3})}}{54(1+\mathcal{C}_{m_2})}\nonumber\\
	   &\approx&\frac{1}{2}(1+\mathcal{C}_{m_2}-\mathcal{C}_{m_1})-\frac{\sqrt{621}}{54}(1-\mathcal{C}_{m_1})\left[1+\frac{1458}{621}\mathcal{C}_{m_1}-\frac{108}{621}(\mathcal{C}_{m_1}+\mathcal{C}_{m_3})\right]^{\frac{1}{2}}\nonumber\\
		&\approx&\frac{1}{2}\left(1-\frac{\sqrt{621}}{27}\right)+\frac{1}{2}(\mathcal{C}_{m_2}-\mathcal{C}_{m_1})+\frac{1}{2\sqrt{621}}\left[27\mathcal{C}_{m_2}-2\mathcal{C}_{m_3}-25\mathcal{C}_{m_1}\right].
		\label{E42}
\end{eqnarray}
The expression $\frac{1}{2}\left(1-\frac{\sqrt{621}}{27}\right)=0.0385...$ denotes Routh criterion. It determines the stability condition for $L_{4,5}$ in classical R3BP. The remaining expressions in the $\mu_c$ equation are due to the effects from the oblateness of the more massive and less massive primaries, radiation pressure of the less and more  massive primaries and PDP.  The critical mass (in $17$ decimal places) can be computed via the substitution of fitting parameters $r_m = 0.99$, $j = 1.5$, $k = 1$, $h = 0.00001$, $c = 1910.83$. Consequently, we find
\begin{eqnarray}
\mu_c&=&0.03086268515886475-0.09804047314470574A_1+0.8293666279034516A_1p_1\nonumber\\
     & &-0.08625060012103034B_1-4.17143645655834750B_1p_1+0.36284935058668233B_2p_2\nonumber\\
		 & &-0.14427151017244513p_1+0.39017449819889993A_2-0.98689512656254470A_2p_1\nonumber\\
		 & &-0.009527626111967606p_2+0.19230243478937234B_2+5.23865406035375200B_2p_1\nonumber\\
		 & &-0.3230170771089508A_1p_2+0.41654634990801520A_2p_2+0.02473745338809197B_1p_2.\nonumber\\
		\label{E43}
\end{eqnarray}
We now proceed to numerically compute the critical mass for the Earth-Moon and Jupiter-Moons systems by using the astrophysical parameters \citep{YE16} presented in Table \ref{Tab2} with $A_1=B_1$ and $A_2=B_2$. Our results are displayed in Table \ref{Tab3}.  The PDP has effect of about $\approx0.01$ reduction on the application of $\mu_c$ to Earth-Moon and Jupiter-Moons systems. The result reflects how ring of dust which could be as a result of asteroid collisions can affect the stability range of an infinitesimal body in R3BP\footnote{For instance, stability range of a spacecraft in Jupiter-Callisto system.}. Furthermore, in figure \ref{fig2}, we have plotted the variation of the critical mass ratio as a function of  $A_2$ and $A_1$ and then varies other parameters in each case. The zonal harmonic coefficients decrease and increase the critical mass in respect to a classical case (i.e. when $B_1 = A_2 = B_2 = p_1 = p_2 = 0$). As it is anticipated, in both figures ($a$ and $b$), it is observe that the critical mass is less than $0.084$.
\begin{table}[ht]
{\caption{Astrophysical parameters for planets systems of our solar system}
\begin{center}
\begin{tabular}{cccc}\\[0.5ex]\hline
S.N.&System&$A_1$&$A_2$\\[2.5ex]\hline
Earth&&&\\[1ex]
1&Moon&0.000000298147&0.000000000000\\[1ex]
Jupiter&&&\\[1ex]
2&Callisto&0.000021186102&-0.000000001213\\[1ex]
3&Metis&0.004586424842&-0.000056862695\\[1ex]
4&Adrastea&0.004515511583&-0.000055117917\\[1ex]
5&Pasiphae&0.000000136024&0.000000000000\\[1ex]
6&Sinope&0.000000133738&0.000000000000\\[1ex]\hline
\end{tabular}\label{Tab2}
\end{center}
\vspace*{-1pt}}
\end{table}
\begin{table}[ht]
{\caption{Critical mass for Earth-Moon and Jupiter-Moons system.}
\begin{center}
\begin{tabular}{ccccc}\\[0.5ex]\hline\hline
\multicolumn{2}{c}{Power-law profile $= 0$}&\multicolumn{1}{c}{}&\multicolumn{2}{c}{Power-law profile $\neq 0$}\\[1.5ex]
{}&{}&{}&{}&{}\\[-1.0ex]\hline
$\mu_c$(Earth-Moon)&$\mu_c$(Jupiter-Callisto)&&$\mu_c$(Earth-Moon)&$\mu_c$(Sun-Callisto)\\[2.5ex]
0.03852079631	&0.03851353085	&&0.03086264505	&0.03080640048\\[1ex]\hline
$\mu_c$(Jupiter-Metis)&$\mu_c$(Jupiter-Adrastea)&&$\mu_c$(Jupiter-Metis)&$\mu_c$(Jupiter-Adrastea)\\[2.5ex]
0.03687828788	&0.03690440886	&&0.02998434157	&0.02994725276\\[1ex]\hline
$\mu_c$(Jupiter-Sinope)&$\mu_c$(Jupiter-Pasiphae)&&$\mu_c$(Jupiter-Sinope)&$\mu_c$(Jupiter-Pasiphae)\\[2.5ex]
0.03852085349	&0.03852085269	&&0.03081027497	&0.03086267535\\[1ex]
\hline
\end{tabular}\label{Tab3}
\end{center}

\vspace*{-1pt}}
\end{table}
\section{Discussions and conclusions}
The results obtain in this study can be reduce to the ones obtained previously if an adjustment is made to the model parameters by setting some of them to zero.  For instance, the equations of motion are in agreement with
\begin{itemize}
\item Classical case of Szebehely \citep{YE19}, if the effects of perturbations and power-law density profile of disc are ignored (i.e. setting $p_1 = p_2 = A_1 = A_2 = B_1 = B_2 =c = 0$)
\item The result of Singh and Taura \citep{YE4} for $M_b=0$ if the effects of power-law density profile of disc (PDP) are ignored (i.e. setting $c = 0$).
\item The result of Singh and Taura \citep{YE20} for $M_b=0$ in the absence of the oblateness up to $J_4$ of the primaries and PDP (i.e. setting $A_2 = B_2 = c = 0$).
\item The result of Kishor and Kushvah \citep{YE13}, if we neglect the effects of the radiation factor of the less massive primary, oblateness coefficients of the more massive primaries and up to $J_4$ of less massive primary (i.e. setting $A_1 = A_2 = B_2 = q_2 = 0$). 
\item Abouelmagd \citep{YE16}, if we consider the oblateness coefficients up to $J_4$ and absence of PDP of the more massive primary (i.e. setting $p_1 = p_2 = B_1 = B_2 = 0$),
\item Singh and Ishwar \citep{YE21}, in the presence of the oblateness up to $J_2$ of the primaries and in absence of PDP (i.e. setting $A_2 = B_2 = c = 0$). 
\item Kushvah \citep{YE17} when only linear terms in small quantities are considered with the effect of the radiation of the more massive primary, oblateness up to $J_2$ of the less massive primary are considered and  ignoring the effect of PDP (i.e. setting $q_2 = 1, A_1 = A_2 = B_2 = c = 0$).
\item Singh and Taura \citep{YE20}, if we disregard the effect of the radiation of the primaries and PDP (i.e., setting $p_1 = p_2 = c = 0$).
\item Jiang and Yeh \citep{YE06} in the absence PDP and all the perturbations (i.e. setting $p_1 = p_2 = A_1 = A_2 = B_1 = B_2 = 0$).
\end{itemize}
In the limiting case where control factor of the density profile ($c$) equals/tend to zero, then, $\mu_c$ obtained in equation (\ref{E43}) becomes
\begin{eqnarray}
\mu_c&=& 0.03852089650455137-0.28500178779055560A_1+0.69578421210336550A_2\nonumber\\
     & &-0.06277956556833330B_1+0.14022865654781014B_2+0.38452273973046225B_2p_2\nonumber\\ 
		 & &-0.00891747059894587p_1+0.62966059506042770A_1p_1-0.4951068898991675A_2p_1\nonumber\\
		 & &-3.75646838678785400B_1p_1+4.71787915998218200B_2p_1-0.00891747059894587p_2\nonumber\\
		 & &-0.31202394234340960A_1p_2+0.41232360442662697A_2p_2-0.00435868211015000B_1p_2\nonumber\\
		\label{E44}
\end{eqnarray}
It is worth mentioning that if we neglect coupling terms $A_ip_i$ and $B_ip_i$ ($i=1,2$) in the above calculation, the result reduces to the one obtained previously in Eq. $(29)$ of ref \citep{YE1} for $M_b=0$ and also in excellent agreement with Eq. $(21)$ of ref \citep{YE20} for $M_b=0$. Furthermore, if we neglect the effect of radiation of the primaries, CT,  and oblateness of the less massive primary (i.e. setting $p_1 = p_2 = B_1 = B_2  = 0$), then equation (\ref{E44}) reduces to the one obtained in ref. \citep{YE16}. Again, if only the oblateness up to $J_2$ of the primaries is considered and ignoring CT (i.e. $A2 = B2 = 0$), then equation (\ref{E44}) reduced to the one obtained in ref.\citep{YE21}. Moreover, by ignoring effect of all  perturbations in equation (\ref{E44}) (i.e. setting $p_1 = p_2 = A_1 = A_2 = B_1 = B_2 = 0$), then our result becomes that of classical CR3BP presented in ref \citep{YE19}

In this paper, we have studied the effect of oblateness up to $J_4$ of the primaries within the framework of restricted three body problem, by using a more realistic model in which a disc with PDP is rotating around the common center of the system mass. The existence and stability of triangular equilibrium points have been explored. It has been demonstrated that triangular equilibrium points are stable for $0<\mu<\mu_c$ and unstable for $\mu_c\leq\mu\leq1/2$, where $\mu_c$ denote the critical mass parameter. Consequently, the oblateness up to $J_2$ of the primaries and the radiation can  reduce the stability range, while the potential from the circular cluster of material points and oblateness up to $J_4$ of the primaries will increase the size of stability both in the context where PDP is considered and ignored. The PDP have effect of about ~0.01 reduction on the application of $\mu_c$ to the Earth-Moon and Jupiter-Moons system. In the limiting case $c=0$, and also by setting appropriate parameter(s) to zero, our results are in excellent agreement with those obtained previously. Our result have a practical application in space dynamics.

Furthermore, the results obtained in ref. \citep{YE11} have been extended and generalized in the current study, by considering the effect of PDP and radiation and radiating factor on the triangular libration points and the critical mass within the framework of R3BP. Our results reflect how ring of dust which could be as a result of asteroid collisions can affect the stability range of an infinitesimal body in R3BP. 

All our numerical computations can be done using a commercial symbolic package such as Mathematica, Maple or MATLAB etc.

Finally, we suggest a possible extension of the current work to a case of non-coplanar equilibrium.

\section*{Acknowledgments}
We thank the kind referees for the positive enlightening comments and suggestions, which have greatly helped us in making improvements to this paper. In addition, BJF acknowledges eJDS (ICTP). OAF acknowledges University of Ilorin for granting him sabbatical leave. The Astrophysics group acknowledges supports from the Vice Chancellor of Federal University Lafia, Prof. Ekanem Ikpi Braide, Dean of Science Faculty, Prof. Emmanuel Kwan-Ndong and the Head of Physics Department, Prof. Sanusi Mohammed Liman. This work is supported partially by 20150964-SIP-IPN, Mexico.

\end{document}